\documentclass{article}
\usepackage{spconf,amsmath,graphicx}

\usepackage[T1]{fontenc} % add special characters (e.g., umlaute)
\usepackage[utf8]{inputenc} % set utf-8 as default input encoding
\usepackage{color}
\usepackage{amsmath,amssymb,graphicx, multirow, booktabs, graphicx, subcaption}
\usepackage{soul}

\title{Multimodal Metric Learning for Tag-based Music Retrieval}
\name{Minz Won$^{\star, \dagger}$ \qquad \thanks{$^{\dagger}$Work performed during an internship with Pandora in 2019.} Sergio Oramas$^{\star\star}$ \qquad Oriol Nieto$^{\star\star}$ \qquad Fabien Gouyon$^{\star\star}$ \qquad Xavier Serra$^{\star}$}
\address{$^{\star}$ Music Technology Group, Universitat Pompeu Fabra, Barcelona, Spain \\
$^{\star\star}$ Pandora, Oakland, CA, United States of America}

\begin{document}
%\ninept
%
\maketitle
\begin{abstract}
Tag-based music retrieval is crucial to browse large-scale music libraries efficiently. Hence, automatic music tagging has been actively explored, mostly as a classification task, which has an inherent limitation: a fixed vocabulary. On the other hand, metric learning enables flexible vocabularies by using pretrained word embeddings as side information. Also, metric learning has already proven its suitability for cross-modal retrieval tasks in other domains (e.g., text-to-image) by jointly learning a multimodal embedding space. In this paper, we investigate three ideas to successfully introduce multimodal metric learning for tag-based music retrieval: elaborate triplet sampling, \textit{acoustic} and \textit{cultural} music information, and domain-specific word embeddings. Our experimental results show that the proposed ideas enhance the retrieval system quantitatively, and qualitatively. Furthermore, we release the \textit{MSD500}, a subset of the Million Song Dataset (MSD) containing 500 cleaned tags, 7 manually annotated tag categories, and user taste profiles.

\end{abstract}
\begin{keywords}
Metric learning, Music retrieval
\end{keywords}

\newcommand{\Minz}{\textcolor{magenta}}

% Introduction
\section{Introduction}\label{sec:introduction}
Text-based search is one of the most common ways of browsing the internet. 
% Text-based search is one of the most common ways of finding information on the web since the internet was introduced. 
This information behavior is also prevalent when exploring music libraries: from querying editorial metadata (e.g., title, artist, album) to high-level music semantics (e.g., genre, mood). However, the annotation of music tags is demanding and time-consuming, especially when an enormous amount of music collections are available. To scale such annotation process, audio-based automatic music tagging has been actively explored by music information retrieval (MIR) researchers~\cite{won2020eval}. 
% Audio-based music tagging is a classification task that predicts whether the given audio content is relevant to a certain music tag based on its acoustic characteristics.
% Thanks to the advent of deep convolutional neural networks (CNN), researchers have proposed ingenious architecture designs with great success in this multi-label binary classification task. 
However, this categorical classification has an intrinsic limitation: it can only use a fixed vocabulary. When an out-of-category tag is queried, music tagging models tend to not properly generalize since the given tag was never considered during training. In a real world scenario, users query a virtually unlimited amount of music tags. Hence, the music retrieval system needs to be more flexible beyond categorical models.

As opposed to categorical classification models, metric learning aims to construct distance metrics for establishing similarity of data~\cite{xing2003distance,weinberger2009distance}. It can form a similarity metric between two instances from the same modality using shared weights, so-called Siamese network~\cite{koch2015siamese}, and this can be also easily expanded towards multiple modalities~\cite{wang2016learning, oramas2018multimodal}.
By jointly learning a multimodal embedding space, metric learning has already demonstrated its suitability for cross-modal retrieval such as image-to-text~\cite{frome2013devise, wang2016learning} and video-to-audio~\cite{suris2018cross}. Metric learning facilitates the nearest neighbor search in the embedding space directly, while classification models require a two-step retrieval (i.e., tagging and ranking). Also, metric learning enables abundant vocabulary when pretrained word embeddings are used to represent tags as side information~\cite{frome2013devise, choi2019zero}. 

Recent work in MIR showed the advantage of using metric learning with pretrained word embeddings for audio-based music tagging and classification~\cite{choi2019zero}. Based on the proposed model, we investigate several ideas to successfully introduce metric learning for tag-based music retrieval.
\noindent\textbf{Contribution. }
Our contribution is four-fold: (\textit{i}) we show the importance of elaborate triplet sampling, (\textit{ii}) we explore \textit{cultural} and \textit{acoustic} information to represent music, (\textit{iii}) we examine domain-specific word embeddings, and (\textit{iv}) we present a manually cleaned dataset for reproducibility.

% This research investigates practical concerns of metric learning to improve the quality of cross-modal music retrieval. The paper is organized as follows. First, we review metric learning and describe our model architectures in Section~\ref{sec:model}. In Section~\ref{sec:dataset}, we present a public dataset, which is a subset of the Million Song Dataset (MSD)~\cite{bertin2011million} with cleaned tags and their categories. Section~\ref{sec:experiments} depicts experimental design to shed light on our practical concerns of metric learning for cross-modal music retrieval: different information sources, domain-specific word embeddings, and an elaborate data sampling method. The following Section~\ref{sec:results} reports experimental results and analyses. Finally, we summarize the conclusion and discuss future work in Section~\ref{sec:discussion}.

% Motivation:
% - Why music retrieval
% - Metric learning in MIR
% - Zero-shot learning paper

% Contribution:
% - Propose a cleaned dataset
% - Facilitate more practical approaches for cross-modal music retrieval (different modalities, word embedding, data sampling, multiple words)
\section{Model}\label{sec:model}
\begin{figure*}[ht!]
    \centering
    \includegraphics[width=0.80\linewidth]{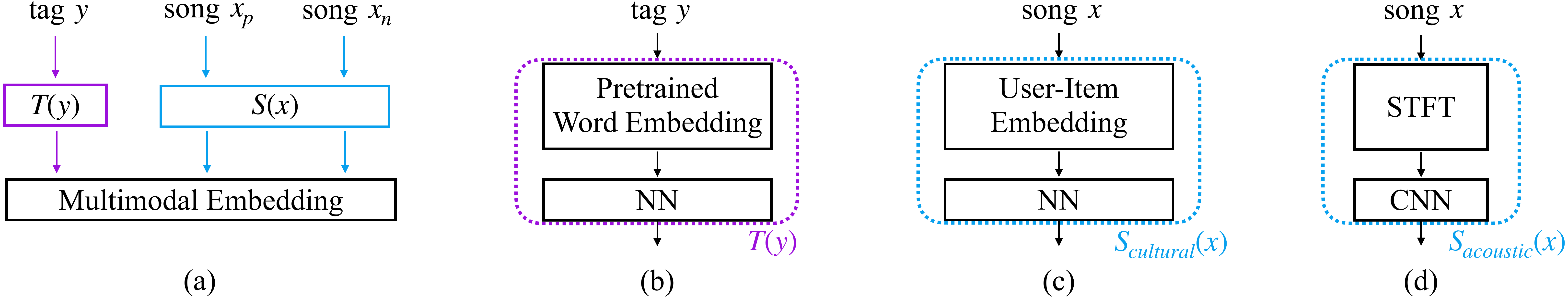} 
    \caption{(a) Overall architecture of the tag-based music retrieval model. (b) Tag embedding branch. (c) Song embedding branch with cultural information. (d) Song embedding branch with acoustic information.}
    \label{fig:model}
\end{figure*}
\subsection{Related work}
% Metric learning aims to learn an appropriate distance metric from labeled examples. The metric learning algorithm learns a transformation of the input space such that the transformed examples in the same class (or in a positive relationship) are brought up together and non-matching examples are far apart in the latent space~\cite{weinberger2009distance}. 
% More specifically, deep metric learning learns nonlinear transformations to measure accurate and robust similarities using deeply stacked neural networks.

A triplet network~\cite{hoffer2015deep} is a type of metric learning that uses a triplet loss to fit a metric embedding, where a positive example $x_p$ belongs to the same class as an anchor $x_a$, and a negative example $x_n$ is a member of a different one. The triplet network is optimized to satisfy $Sim(x_a, x_p) > Sim(x_a, x_n)$, where $Sim(.)$ is a learned similarity metric. As it learns by comparisons instead of using direct labels, the triplet approach is expandable to leverage various data sources that are not explicitly labeled. Thanks to its flexibility, deep metric learning with the triplet loss has been actively used to solve a set of diverse MIR problems~\cite{oramas2018multimodal,choi2019zero}.

% Park et al. \cite{park2017representation} obtain a versatile music representation that can be used for similarity-based music retrieval and genre classification. Artist labels are objective hence they are less noisy and less expensive than semantic labels. Authors regard two different audio clips to be in a positive relationship if they are from the same artist, and the negative relationships are defined otherwise. Since it shares the same embedding function $f(x)$ for original, positive, and negative embeddings, this is also called a Siamese network~\cite{koch2015siamese}.

% Oramas et al. \cite{oramas2018multimodal} employ deep metric learning to form a multimodal space with album cover artworks and music audio. A learnable audio embedding function $f(x)$ maps the input audio to the shared embedding and an image embedding function $g(z)$ learns to map the input image $z$ to the shared embedding space, thus obtaining a two-branch neural network \cite{wang2016learning}. The pretrained multimodal embedding is later used for music genre classification successfully. 

Choi et al. \cite{choi2019zero} proposed a triplet network that learns a multimodal embedding of audio and word semantics. To handle unseen labels, the authors used pretrained GloVe embeddings \cite{pennington2014glove} as side information.
% for their word semantic inputs. 
An audio embedding branch learns the mapping of the audio input to the multimodal embedding space. And another branch maps pretrained word embeddings to the shared multimodal embedding space. This metric learning model with side information demonstrated its versatility in multi-label zero-shot annotation and retrieval tasks. Since it can perform cross-modal retrieval (i.e., text-to-music), we adopt this architecture design as the backbone of our tag-based music retrieval model.
% An audio embedding function $f(x)$ learns the mapping of the audio input to the multimodal embedding space. And another branch can be described as $h(w(y))$ where $w(y)$ is the pretrained word embedding of tag $y$ and $h(.)$ is a learnable mapping function of tag semantics. This metric learning model with side information demonstrated the versatility of the pretrained embedding in a multi-label zero-shot annotation and a multi-label zero-shot retrieval tasks. Since it can perform cross-modal retrieval (i.e., retrieve audio using a tag input or retrieve tags using an audio input), we adopt this architecture design as the backbone of our tag-based music retrieval model.

\subsection{Model description}
\subsubsection{Architecture overview}
% In this paper, we implement a two-branch neural network~\cite{wang2016learning} that resembles the work in~\cite{choi2019zero}. 
Similar to the previous work~\cite{choi2019zero}, our model comprises of two branches. One branch $T(y)$ learns the mapping of tag semantics $y$ to the embedding space, and another branch $S(x)$ learns the mapping of song information $x$ to the shared embedding space --- see Figure~\ref{fig:model}-(a). The model is trained to minimize the following loss function $L$:
\begin{equation}
L = [D(E_a, E_p) - D(E_a, E_{n}) + \delta]_+,
\end{equation}
where $D$ is a cosine distance function, $\delta$ is a margin, and $E_a$, $E_p$, $E_n$ are mapped embeddings of anchor tag, positive song, and negative song, respectively. $[\cdot]_+$ is a rectified linear unit (ReLU). The margin $\delta$ prevents the network from mapping all the embeddings to be the same (i.e., $L=0$ for any inputs). With learnable transformations $T(y)$ and $S(x)$, the equation can be rewritten as:
\begin{equation}
L = [D(T(y_a), S(x_{p})) - D(T(y_a), S(x_{n})) + \delta]_+,
\end{equation}
where $y_a$ is the anchor tag input, and $x_p$ and $x_n$ are positive and negative song inputs, respectively. Following subsections depict the details of each branch $T(y)$ and $S(x)$.

\subsubsection{Tag embedding}
Figure~\ref{fig:model}-(b) shows the tag embedding branch $T(y)$. A given tag $y$ passes through the pretrained word embedding model which results in a 300-dimensional vector. By using the pretrained word embeddings, the system can handle richer vocabulary than categorical models. For example, one can expect the system to handle plural forms (\textit{guitar} and \textit{guitars}), synonyms (\textit{happy} and \textit{cheerful}), acronyms (\textit{edm} and \textit{electronic dance music}), and dialectal forms (\textit{brazil} and \textit{brasil}). As our baseline, we use Word2Vec~\cite{mikolov2013efficient} embeddings pretrained with Google News dataset. The tag embedding is input to a neural network which is fully connected to a 512-dimensional hidden layer followed by a 256-dimensional output layer.

% Figure~\ref{fig:model}-(b) shows the tag embedding branch $T(y)$. A given tag $y$ passes through the pretrained word embedding $w(.)$ which results in a 300-dimensional vector. By using the pretrained word embedding, the system can handle richer vocabulary than categorical models. For example, one can expect the system to handle plural forms (e.g., \textit{guitar} and \textit{guitars}), synonyms (e.g., \textit{happy} and \textit{cheerful}), acronyms (e.g., \textit{edm} and \textit{electronic dance music}), and foreign language forms (e.g., \textit{brazil} and \textit{brasil}). As our baseline, we use Word2Vec~\cite{goldberg2014word2vec} embedding pretrained with Google News dataset. The tag embedding is input to a neural network which is fully connected to a 512-dimensional hidden layer followed by a 256-dimensional output layer.

\subsubsection{Song embedding}
Pachet et al.~\cite{pachet2005knowledge} outlined three main types of music information: \textit{editorial}, \textit{cultural}, and \textit{acoustic}. Most of the previous works in music tagging~\cite{won2020eval} and multimodal metric learning \cite{oramas2018multimodal, choi2019zero}, focused mainly on acoustic information to represent music. In our work, we attempt to harness not only acoustic information but also cultural information in music retrieval. Cultural information is produced by the environment or culture. One of the most common methods to obtain cultural information is collaborative filtering~\cite{herlocker2000explaining}. 
% \st{We employ the Echo Nest Taste Profile Subset \cite{McFee2012} to process collaborative filtering.}\ON{I would not talk about the dataset here}

The song embedding branch with cultural information $S_{cultural}(x)$ consists of a user-item embedding and a neural network --- Figure~\ref{fig:model}-(c). The user-item embedding is obtained by factorizing a user-song interaction matrix.  %using the MSD Taste Profile~\cite{bertin2011million}. The matrix of user-item interactions is factorized using 
Weighted matrix factorization (WMF) with the alternating least squares (ALS)~\cite{Hu2008} is used, yielding both user and song embeddings of 200 dimensions each. User embeddings are discarded and song embeddings are used as our input. The input of the neural network is fully connected to a 512-dimensional hidden layer followed by a 256-dimensional output layer.

The song embedding branch with acoustic information $S_{acoustic}(x)$ learns audio-based music representation using a convolutional neural network (CNN) --- Figure~\ref{fig:model}-(d). According to previous research~\cite{won2020eval}, a simple 2D CNN with $3\times3$ filters could achieve competitive results to state-of-the-art in music tagging when it uses a short chunk of audio inputs ($\approx$4s). For simplicity, we adopt the short-chunk CNN~\cite{won2020eval} to train our acoustic embedding.
% instead of other complex state-of-the-art models~\cite{won2020data, won2020eval}. Short-chunk CNN consists of seven convolutional layers. % Each layer comprises convolution, batch normalization~\cite{ioffe2015batch}, ReLU, and max-pooling. 

The model is optimized using ADAM~\cite{kingma2014adam} with $10^{-4}$ learning rate, and $10^{-4}$ weight decay. The model is trained for 200 epochs where 1 epoch includes 10,000 triplets. 
% We select the best models based on validation loss after 300 epochs of training where 1 epoch includes 5,000 triplets.
For input preprocessing, audio files are downsampled to 22.5kHz then converted to Mel spectrograms using 1024-point FFT with 50\% overlap and 128 Mel bands.

\section{Dataset}\label{sec:dataset}
The Million Song Dataset (MSD) \cite{bertin2011million} is a collection of metadata and precomputed audio features for 1 million songs. Along with this dataset, the Echo Nest Taste Profile Subset \cite{McFee2012} provides play counts of 1 million users on more than 380,000 songs from the MSD, and the Last.fm Subset provides tag annotations to more than 500,000 songs from the MSD. We take advantage of these two subsets of the MSD to build our own dataset. Tags in the Last.fm Subset are very noisy, including 522,366 distinct tags. We performed a cleanup process of the dataset (e.g., merge synonyms or acronyms, fix misspelling) in order to have fewer tags while supported with a reasonable number of annotations. The detailed cleanup process is described in our online repository.

% The cleanup process consists of the following steps:
% \begin{itemize}
%     \item Filter out all tracks not included in the MSD Taste Profile.
%     \item Filter out all tag annotations with a Last.fm tag score of 0 (Last.fm tags in the original dataset come with a score between 0 and 100).
%     \item Filter out all tracks with more than 20 tags (we assume that annotations in tracks with too many tags are less reliable).
%     \item Preprocessing of tags: remove punctuation, normalize expressions (e.g., and, \&, 'n'), and remove irrelevant suffixes (e.g., music, song, tag).
%     \item Group all tags with the same preprocessed lemma and name the group with the most common of the involved tags.
%     \item Select all tag groups with annotations for at least 100 tracks.
% \end{itemize}

The final dataset contains 500 tag groups (from now on we will simply call these groups ``tags''), which subsumes 1,352 distinct Last.fm tags. These 500 tags are then manually classified in a lightweight taxonomy of 7 classes (genre, mood, location, language, instrument, activity, and decade). 158,323 distinct tracks are tagged with these 500 tags with an average of 3.1 tags per track and each track has user play counts. We release the final dataset as the \textit{MSD500}.

% \begin{table}[h]
% \centering
% \begin{tabular}{c|c}
%     \textbf{Class} & \textbf{Number of tags}  \\
%     \hline
%     genre & 294 \\
%     mood/character & 94 \\
%     location & 36 \\
%     language/origin & 34 \\
%     instrument & 21 \\
%     activity & 14 \\
%     decade & 7 \\
% \end{tabular}
% \caption{\textit{MSD500} number of tags per class}
% \label{tbl_dataset}
% \end{table}

In this paper, we use two different subsets of the proposed dataset which are \textit{MSD100} and \textit{MSD50}. 
Music tags are highly skewed towards few popular tags and handling this skewness is another big topic in data-driven approaches. Models are optimized to predict more popular tags in the training set while evaluation metrics are averaged over tags. To avoid the undesired effect of the high skewness, we only use the top 100 tags in our experiments which results in 115k songs (\textit{MSD100}). 

Although we have user information in our dataset, the interaction counts are not scalable compared to industry standards~\cite{korzeniowski2020mood}. This can possibly mislead us to overlook the representation power of cultural information. Hence, we build another subset which includes 39,402 songs with Last.fm tags and user-item embeddings from more than 100B in-house user explicit feedback. In this case we only use top 50 tags (\textit{MSD50}) because the size of the dataset became smaller during the mapping process. As the in-house user feedback includes sensitive information, we only release the song IDs and their tags of the \textit{MSD50}. All data splits have been done in artist-level to avoid unintentional information leakage.
\section{Experiments}\label{sec:experiments}
% In this section, we introduce three experiments which can be critical for improving the quality of metric learning for tag-based music retrieval. 
In this section, we introduce three experiments which can be critical to enhance our metric learning approach for tag-based music retrieval. 
All models are evaluated with mean average precision (MAP) over the labels and precision at 10 (P@10).
% , and area under the receiver operating characteristic curve (ROC-AUC) averaged over the label. 
% Note that since our main interest is music retrieval, not music tagging, MAP and P@10 are our main metrics. 
Reproducible code and dataset are available online~\footnote{https://github.com/minzwon/tag-based-music-retrieval}.

\begin{table}[t!]
\centering
\footnotesize
\begin{tabular}{@{\hskip 0.12in}l@{\hskip 0.32in}c@{\hskip 0.32in}c@{\hskip 0.32in}c@{\hskip 0.12in}}
\toprule
Metrics & Random & Balanced & Balanced-weighted\\ \midrule
MAP& 0.1658 & 0.1675 & \textbf{0.1852}\\
P@10 &  0.2990  & 0.3160 & \textbf{0.3500}\\ 
% ROC-AUC& 0.8588 & 0.8602 & \textbf{0.8656}\\
 \bottomrule
\end{tabular}
\caption{Performance of different sampling methods.}
\label{tab:sampling}
\end{table}

\subsection{Sampling matters}
% The number of possible triplets grows cubically as the number of observations grows~\cite{wang2014learning}.
The number of possible triplets grows cubically as the number of observations grows. Thus, triplet sampling is crucial in deep metric learning~\cite{wu2017sampling}, as it matters equally or more than the choice of loss functions. In this subsection, we explore three different sampling methods: random sampling, balanced sampling, and balanced-weighted sampling.

Random sampling randomly chooses one song to generate an anchor-positive pair. Then a negative example is randomly sampled from a set of songs without the anchor tag. With this method, more popular tags are more likely to be sampled as the anchor tag. Also, songs with minor tags are less likely to be sampled as negative examples. 

To alleviate this problem, the balanced sampling method uniformly samples an anchor tag first and then select a positive song. Minor tags may have equal possibilities to popular tags to be sampled as an anchor tag. By sampling negative examples from the batch of the positive songs, we can also expect more balanced tag distribution of negative examples. 

For more efficient training, various triplet sampling methods have been proposed such as hard negative mining~\cite{simo2015discriminative}, semi-hard negative mining~\cite{schroff2015facenet}, and distance weighted sampling~\cite{wu2017sampling}. We combine the distance weighted sampling~\cite{wu2017sampling} with the aforementioned tag balancing method (i.e., balanced-weighted sampling). Identical to the balanced sampling, we select an anchor tag and a positive song. From the given batch of positive songs, we sample negative examples based on their cosine distances from the anchor tags in the embedding space. Thus, more informative (harder) negative examples are more likely to be sampled while not loosing semi-hard and soft negative examples.

As shown in Table~\ref{tab:sampling}, balanced-weighted sampling outperforms other sampling methods. This proves that sampling matters for training our tag-based music retrieval model. Note that here we only used acoustic information for the song embedding to control the experiment. From now on, the following experiments use the balanced-weighted sampling method.

\subsection{Acoustic and cultural music representation}
We believe certain groups of tags are more related to acoustic information while others may be more culturally relevant. A tag \textit{piano}, for example, can be predicted using the user-item matrix if there is a specific group of users who heavily listened to songs with piano. However, originally, the tag \textit{piano} is associated with acoustic information. When there is a song beloved by the aforementioned user group, if we only use cultural information, the song can be regarded as piano music even when no piano can be acoustically perceived in the song. 
As another example, a tag \textit{K-pop} can be predicted based on acoustic information since there are common acoustic characteristics of \textit{K-pop}. However, if the song is not from Korea and is not being consumed in Korea, it should not be tagged as \textit{K-pop}. 
To investigate the capability of two different information sources, we train our metric learning model with cultural information only and acoustic information only: $S_{cultural}$ and $S_{acoustic}$, respectively.

\begin{figure}
 \centerline{
 \includegraphics[width=0.97\columnwidth, trim=6 4 4 6, clip]{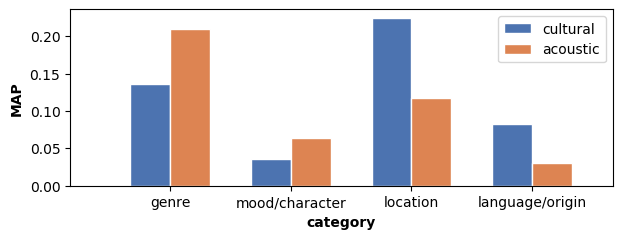}}
 \caption{Category-wise MAP of cultural and acoustic models on \textit{MSD100}.}
 \label{fig:tagwise}
\end{figure}

As shown in Table~\ref{tab:multimodal}, the acoustic model outperforms the cultural model in overall metrics with \textit{MSD100}. However, if we take a closer look at category-wise scores, the cultural model shows its strength in \textit{location} and \textit{language/origin} tags (Figure~\ref{fig:tagwise}). This supports our hypothesis that the modality selection has to be associated with its original source of information. But more important factor than the information source is the size and quality of available data. In Table~\ref{tab:multimodal} (\textit{MSD50}), as cultural information becomes richer (Cul-I), the cultural model outperforms the acoustic model. In addition, we observed that the cultural model with richer information (Cul-I) is superior in every tag category including \textit{genre} and \textit{mood}. As observed, acoustic and cultural models show different strengths, but the foremost important factor of the modality selection is the size and quality of available user-item interactions and audio data. We also experimented a hybrid model with simple concatenation of cultural and acoustic embeddings but it did not improve (Table~\ref{tab:multimodal}-Concat).

\begin{table}[t!]
\centering
\footnotesize
\begin{tabular}{@{}c|ccc|ccc@{}}
\toprule
\multicolumn{1}{c}{\multirow{2}{*}{Metrics}} & \multicolumn{3}{c}{MSD100} & \multicolumn{3}{c}{MSD50} \\ \cmidrule(l){2-7} 
% \multicolumn{1}{c}{} & Cultural-E & Acoustic & Concat & Cultural-E & Cultural-P & Acoustic \\ \midrule
\multicolumn{1}{c}{} & Cul-E & Acoustic & Concat & Cul-E & Cul-I & Acoustic \\ \midrule
MAP & 0.1155 & \textbf{0.1852} & 0.1775 & 0.2163 & \textbf{0.4719}  & 0.3062\\
P@10   & 0.3200  & \textbf{0.3500} & 0.3120 & 0.4500 & \textbf{0.6380} & 0.4680                  \\
% ROC-AUC     & 0.7674 & \textbf{0.8656} & 0.8617 & 0.7933 & \textbf{0.9243} & 0.8863   \\ 
\bottomrule
\end{tabular}
\caption{Performance of cultural and acoustic models. Cul-E and -I use the EchoNest Taste Profiles and our in-house user explicit feedbacks, respectively.}
\label{tab:multimodal}
\end{table}

% \begin{table}[t!]
% \centering
% \footnotesize
% \begin{tabular}{@{\hskip 0.12in}l@{\hskip 0.32in}c@{\hskip 0.32in}c@{\hskip 0.32in}c@{\hskip 0.12in}}
% \toprule
% Metrics & Random & Balanced & Balanced-weighted\\ \midrule
% MAP& 0.1658 & 0.1675 & \textbf{0.1852}\\
% P@10 &  0.2990  & 0.3160 & \textbf{0.3500}\\ 
% ROC-AUC& 0.8588 & 0.8602 & \textbf{0.8656}\\
%  \bottomrule
% \end{tabular}
% \caption{Performance of different sampling methods.}
% \label{tab:sampling}
% \end{table}

\begin{table}
\centering
\footnotesize
\begin{tabular}{|@{\hskip0.8pt}c@{\hskip0.8pt}|@{\hskip0.8pt}c@{\hskip0.8pt}|@{\hskip0.8pt}c@{\hskip0.8pt}|}
\hline
Tag    & GoogleNews                                                                                                                                                                                        & Domain-specific                                                                                                                                             \\ \hline
Jungle & \begin{tabular}[c]{@{}c@{}}jungles, dense\_jungle, \\ dense\_jungles, rainforest, \\ thick\_jungles, Amazon\_jungle, \\ Amazonian\_jungle, steamy\_jungles, \\ hilly\_jungle, swamps\end{tabular} & \textbf{\begin{tabular}[c]{@{}c@{}}breakbeat, dub, drum\_n\_bass, \\ drum'n'bass, grime, \\ deep\_house, ragga, dubstep, \\ acid, acid\_house\end{tabular}} \\ \hline
\end{tabular}
 \caption{Nearest words in GoogleNews and domain-specific word embeddings. Music-related words are emboldened.}
 \label{tab:neighbor}
\end{table}

\subsection{Domain-specific word embeddings}
We use pretrained Word2Vec~\cite{mikolov2013efficient} embeddings as a part of our tag branch $T(y)$. Since the word embeddings are trained with Google News, it is hard to expect the trained embeddings to have musical context. 
% For example, when we look up the definition of \textit{house} from the dictionary \cite{webster2006merriam}, its first definition is: \textit{a building that serves as living quarters for one or a few families}. However, in a musical context, the tag \textit{house} means the ninth definition on the dictionary which is: \textit{a type of dance music mixed by a disc jockey that features overdubbing with a heavy repetitive drumbeat and repeated electronic melody lines}. To investigate the impact of a domain-specific word embedding which contains more musical context, we pretrain our own word embedding with more musical text data.

We pretrain our own word embeddings with musical text data. We use the corpus of text from the subtask 2B of the SemEval-2018 Hypernym Discovery Task~\footnote{https://competitions.codalab.org/competitions/17119\#learn\_the\_details-terms\_and\_conditions}. It contains an already tokenized 100M-word corpus including Amazon reviews, music biographies, and Wikipedia pages about theory and music genres. We train a Word2Vec model on this corpus with a window of 10 words yielding word embeddings for unigrams, frequent bigrams and trigrams of 300 dimensions. 

We could not discover any quantitative performance gain by using of our domain-specific word embeddings. However, as shown in Table~\ref{tab:neighbor}, the domain-specific word embeddings could include more musical context in it. For example, for the unseen query \textit{jungle}, a model with domain-specific embeddings could successfully retrieve relevant items while conventional embeddings could not. Also, domain-specific music corpora include frequent bigrams and trigrams, such as \textit{deep house} or \textit{smooth jazz}, which are not typically captured in word embeddings trained on general text corpora. More qualitative examples are included in our online repository.

\section{Conclusion}\label{sec:conclusion}
In this paper, we explored three different ideas to enhance the quality of metric learning for tag-based music retrieval. Balanced-weighted sampling could successfully improve the evaluation metrics. Cultural and acoustic models showed different strengths based on the information source of the given tag but the foremost important factor is the size and quality of available data. Finally, domain-specific word embeddings showed its suitability for music retrieval by including more musical context.

As future work, in-depth comparison of acoustic and cultural models is necessary to better understand how the size and the quality of data affect the results. Also, a hybrid method of fusing acoustic and cultural information has to be explored as simple concatenation did not bring any improvement. In addition, further evaluation of out-of-vocabulary tags is needed to determine the real impact of domain-specific word embeddings. Finally, to meet the real-world demand, multi-tag retrieval systems have to be considered.

% For future work, we would like to explore three directions: each direction is related to song embeddings, tag embeddings, and triplet sampling, respectively. As reported in this paper, hybrid multimodal song embeddings improve performance but not for every single tag. Therefore, further analyses are required to effectively merge different sources of information. Current tag embeddings only enable single-tag queries while most of the songs in the real world have multiple tags. To facilitate more flexible retrieval systems, multi-tag processing has to be considered. 
\section{Acknowledgement}\label{sec:acknowledgement}
This work was funded by the predoctoral grant MDM-2015-0502-17-2 from the Spanish Ministry of Economy and Competitiveness linked to the Maria de Maeztu Units of Excellence Programme (MDM-2015-0502).

\bibliographystyle{IEEEbib}
\bibliography{strings,refs}

\end{document}